\documentclass[aip,apl,reprint]{revtex4-1}  
\usepackage[dvips,final]{graphics}
\usepackage{graphicx}
\usepackage{amsfonts}              
\usepackage{amssymb,amsmath}
\usepackage{bm}
\usepackage{array}
\usepackage{natbib}
\usepackage{comment}
\begin{document}
\title{Topological Insulator Cell for Memory and Magnetic Sensor Applications}
 \author{T.\ Fujita}       
   \affiliation{Computational Nanoelectronics and Nano-device Laboratory, Electrical and Computer Engineering Department, National University of Singapore, 4 Engineering Drive 3, Singapore 117576}
   \author{M.\ B.\ A.\ Jalil}       
   \email{elembaj@nus.edu.sg}
   \affiliation{Information Storage Materials Laboratory, Electrical and Computer Engineering Department, National University of Singapore, 4 Engineering Drive 3, Singapore 117576}
   \affiliation{Computational Nanoelectronics and Nano-device Laboratory, Electrical and Computer Engineering Department, National University of Singapore, 4 Engineering Drive 3, Singapore 117576}
    \author{S.\ G.\ Tan}       
    \affiliation{Data Storage Institute, A*STAR (Agency for Science, Technology and Research) DSI Building, 5 Engineering Drive 1, Singapore 117608}
   \affiliation{Computational Nanoelectronics and Nano-device Laboratory, Electrical and Computer Engineering Department, National University of Singapore, 4 Engineering Drive 3, Singapore 117576}
    \date{\today}    
\begin{abstract}
We propose a memory device based on magnetically doped surfaces of 3D topological insulators. Magnetic information stored on the surface is read out via the quantized Hall effect, which is characterized by a topological invariant. Consequently, the readout process is insensitive to disorder, variations in device geometry, and imperfections in the writing process.
\end{abstract}
\maketitle
Integer quantum numbers in physics can arise either from symmetry, for example the quantized angular momentum associated with rotational symmetry, or from topological considerations, such as the quantized Hall conductivity in the integer quantum Hall effect (IQHE).\cite{thouless} The key distinction between these two scenarios is their robustness to perturbations; discreteness in the former breaks down in the presence of perturbations which remove the symmetry, whilst in the latter the quantum numbers remain preserved even under relatively strong perturbations (such as disorder, system geometry and so forth). Such quantum numbers which are protected by topology are called \emph{topological invariants}. From a practical standpoint, it is of interest to measure and make use of such quantities. The most notable experimental measurement of a topological invariant is the quantized Hall conductivity of the IQHE in semiconductor quantum wells.\cite{klitzing} Owing to its remarkable precision, this is used as a measure for the international standard for electrical resistance. On the other hand, topological invariants might also be attractive in device applications with highly robust characteristics. 

In this letter, we examine a new form of non-volatile magnetic storage, in which the writing process entails a conventional writing field, but whose electrical readout process is topologically protected and is consequently robust against weak disorder and perturbations. The basis for our proposed device is the Hall effect mediated by the $\vec{k}$-space Berry curvature in the presence of spin-orbit coupling (SOC). In a recent paper, Qi \emph{et al}.\cite{qi}\ proposed a general 2D model for the quantum anomalous Hall effect for general SOC systems, and found that the charge Hall conductivity is topologically quantized in analogy with the IQHE. Here, we study a practical realization of this system which should exist on the metallic surfaces of 3D topological insulators (TIs). In particular we devise a TI-based magnetic memory cell, in which a bit is stored via the exchange coupling of the TI surface states (SS) induced by magnetic doping. The magnetism induces a finite $\vec{k}$-space Berry curvature in the SS, thereby driving the Hall effect.\cite{jzhang} The readout (Hall) voltage of the cell is related directly to the Hall conductivity, which is highly sensitive to the magnetization of the FM film (\emph{i.e.}\ the stored bit) but which is insensitive to weak disorder, cell imperfections, and cell geometry.

\begin{figure}[!ht]
\centering
\resizebox{\columnwidth}{!}{
\includegraphics{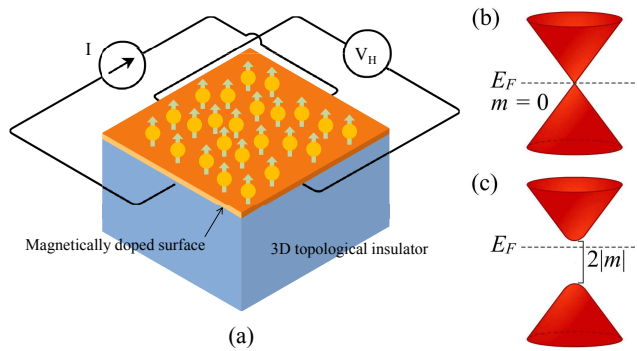}}
\caption{(Color online) (a) Structure of proposed memory cell, based on a topological insulator (TI) block with a magnetically doped surface. A bit is stored by the perpendicular magnetization of the surface. The readout process is facilitated by the Hall effect; a source drives a current through the TI surface, and Hall electrodes measure the resulting Hall voltage $V_H$. (b) Energy band structure of neutral TI surface states (SS), showing the conduction and valence bands. (c) Massive, gapped energy band structure corresponding to magnetically doped SS.}
\label{device.fig}
\end{figure}

\emph{Device structure.} We propose the memory cell illustrated in Fig.\ \ref{device.fig}(a), based on a 3D TI block (\emph{e.g.}\ Bi$_2$Te$_3$, Bi$_2$Sb$_3$, Sb$_2$Te$_3$). The SS of neutral TIs comprise of a single Dirac cone centered at the $\Gamma$-point, which have been observed by ARPES measurements,\cite{hsieh,xia} with the Fermi energy lying at the Dirac point; see Fig.\ \ref{device.fig}(b). Time-reversal (TR) symmetry  guarantees degeneracy of the states at $\vec{k}=\Gamma$, \emph{i.e.}\ the surface states are gapless and are protected by Kramer's theorem.\cite{kane} We consider ferromagnetic (FM) doping of the TI surface which induces long range magnetic order.\cite{biswas} This breaks TR symmetry and opens up an energy gap at the Dirac point. Assuming that the magnetic surface has perpendicular magnetic anisotropy (for discussion, see Refs.\ [\onlinecite{biswas}], [\onlinecite{wray}]), the effective Hamiltonian of the two-dimensional SS is
\begin{equation}
\mathcal{H}=v_F\vec{\sigma}\cdot\left(\vec{p}\times\hat{z}\right)+m\sigma^z,\label{H.eq}
\end{equation}
where $\vec{p}=\left(p_x,p_y\right)=\hbar\vec{k}$ is the in-plane momentum, $v_F$ is the Fermi velocity, $m$ is the internal exchange splitting, $\vec{\sigma}=\left( \sigma^x,\sigma^y\right)$ and $\sigma^i$ ($i=x,y,z$) are the Pauli spin matrices. The energy eigenvalues are $\mathcal{H}|\psi_\tau\rangle =E_\tau|\psi_\tau\rangle=\tau\sqrt{v_F^2{p}^2+m^2}|\psi_\tau\rangle$, where $|\psi_\tau\rangle$ are the single particle eigenstates, and $\tau=\pm 1$. Thus, the energy spectrum comprises of a conduction band (CB; $\tau=+1$) and valence band (VB; $\tau=-1$), separated by a gap $\Delta=2|m|$ at $\vec{k}=0$, as illustrated in Fig.\ \ref{device.fig}(c). Preliminary experiments on FM doping of TI surfaces reveal an energy gap of order $\Delta\sim10$ meV ($\sim 100$ K) for modestly doped samples ($1\%$ Mn-doped Bi$_2$Se$_3$).\cite{ylchen} Alternatively, the TR of the SS can be broken by a FM film coating the TI surface.\cite{jzhang,nomura,yokoyama} In this case, an insulating FM film is required (\emph{e.g.}\ EuO or EuS \cite{santos}) as the transport should remain solely at the TI surface. One caveat of this approach, however, is the relatively small gap size of $\Delta\sim 1$ meV ($\sim 10$ K) induced by FM films.\cite{jzhang} On the other hand, this technique boasts the advantage of FM films having relatively high Curie temperatures $T_C$ (see discussions regarding $T_C$ later).

\emph{Berry curvature and Hall effect physics.} 
In addition to the energy gap, the broken TR symmetry induces a finite Berry curvature in the crystal momentum $\vec{k}$-space. The appearance of the Berry curvature is ubiquitous in physics, and is relevant in many contexts \emph{e.g.}\ in spintronics, optics and graphene. Carriers that are subject to the $\vec{k}$-space Berry curvature undergo anomalous transport,\cite{kl} which drives the Hall effect. Each eigenband is endowed with a Berry connection, defined by $\mathcal{A}_\tau(\vec{k})\equiv -i\langle\psi_\tau|\nabla_{\vec{k}}|\psi_\tau\rangle$, where $|\psi_\tau\rangle$ are the single particle eigenstates of the Hamiltonian in Eq.\ \eqref{H.eq}. This quantity represents a vector potential in crystal $\vec{k}$-space. The corresponding Berry curvature $\Omega_k=\epsilon_{ijk}\left(\partial_i\mathcal{A}_j-\partial_j\mathcal{A}_i\right)$ then represents an effective $\vec{k}$-space magnetic field, 
\begin{equation}
\Omega_z(\vec{k})=\tau\frac{mv_F^2\hbar^2}{2\left( m^2+v_F^2\hbar^2(k_x^2+k_y^2)\right)^{\frac{3}{2}}}.
\end{equation}
The Hall conductivity of the SS can be quantified via the Kubo formula, which reveals its direct relation to the Berry curvature, \emph{i.e.}\ $\sigma_{xy}=\left(e^2/\hbar\right)\int \frac{d^2\vec{k}}{(2\pi)^2}(f_+-f_-)(\vec{k})\Omega_z(\vec{k})$,\cite{qi,lu}
where $f_\tau(\vec{k})=1/\left[\exp{\left((E_\tau(\vec{k})-E_F)/k_BT\right)}+1\right]$ is the Fermi distribution of band $\tau=\pm$, $k_B T$ is thermal energy, and the integration is carried out over all occupied states.  The limits of integration in the expression for $\sigma_{xy}$ are governed by the position of the Fermi level $E_F$ with respect to the bands. When the Fermi level $E_F$ lies inside the energy gap [see Fig.\ \ref{device.fig}(c)] so that the VB is completely full and the CB completely empty ($f_{-}=1$, $f_+=0$), the Hall conductivity in the low temperature limit takes on the half-quantized value
\begin{equation}
\sigma_{xy}=\frac{e^2}{\hbar}\int_0^\infty \frac{d^2\vec{k}}{(2\pi)^2}\Omega_z(\vec{k})=-\frac{e^2}{2h}\text{sgn}(m)\label{winding.eq}
\end{equation}
which is a topological invariant.\cite{qi,jzhang} This situation is reminiscent of the perfectly quantized Hall conductivity in the integer quantum-Hall effect (IQHE). Here, $\sigma_{xy}$ is finite despite an insulating bulk state due to the presence of conducting edge channels, just as in the IQHE. 

\begin{figure}[!ht]
\centering
\resizebox{\columnwidth}{!}{
\includegraphics{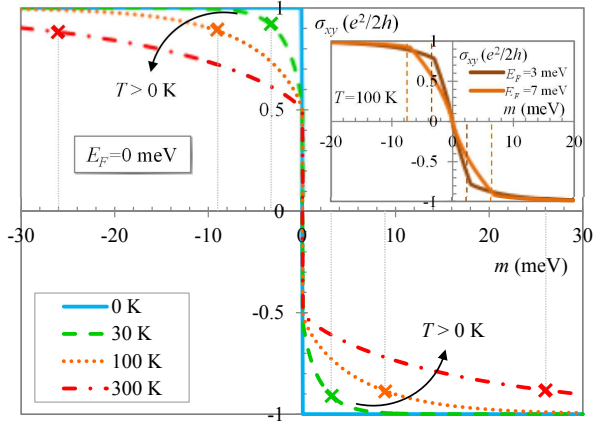}}
\caption{(Color online) (main) Hall conductivity $\sigma_{xy}$ as a function of $m$ for various temperatures $T$, with the Fermi level fixed at $E_F=0$ meV. When $T=0$ K, $\sigma_{xy}=\pm e^2/2h$ is perfectly quantized \eqref{winding.eq}. This facilitates a topological readout process of a magnetic bit represented by the sign of $m$. For finite temperatures, $\sigma_{xy}$ is diminished due to scaling by the Fermi distribution; however, the accuracy is restored in the large $|m|$ limit. For $T>0$ K, we require a sufficiently large gap $\Delta=2|m|\gg k_B T$ separating the conduction and valence bands. For illustration, we mark the points $|m|= k_B T$ for each $T$. (inset) $\sigma_{xy}$ as a function of $m$ for $T=100$ K, and varying $E_F$. For our cell operation, we require $|m|>E_F$, where $\sigma_{xy}$ is fairly robust with value $\sigma_{xy}\approx \pm e^2/2h$ which improves with large $|m|$.}
\label{hall_cond_topological.fig}
\end{figure}

\emph{Reading and writing to memory cell.}
In our memory cell, a bit is stored by the magnetization $\vec{M}$ of the FM-doped TI surface, with, say, a $``1"$ ($``0"$) being stored by an upward (downward) pointing $\vec{M}$. Writing to the cell would require a writing field whose field strength exceeds the magnetic coercivity of the surface. Previous works have indicated that magnetically doped Bi$_2$Te$_3$ should have a coercivity of $H_C\sim 0.01$ T,\cite{garate} which has been measured by experiment (Mn-doped Bi$_2$Te$_3$).\cite{yshor} The soft magnetic anisotropy may be viable in MRAM and magnetic sensor applications, as it reduces the required switching field.  The stored bit is read out from the cell via the Hall voltage $V_H$ which is inversely proportional to $\sigma_{xy}$,
\begin{equation}
V_H=\frac{I}{\sigma_{xy}},
\end{equation}
 where $I$ is a constant current flowing through the device.\footnote{This equation applies only when the Fermi level lies in the gap and the longitudinal conductivity $\sigma_{xx}$ vanishes} A current source provides an electric current $I$ across the surface, whilst Hall electrodes are attached to the lateral sides to measure $V_H$ as depicted in Fig.\ \ref{device.fig}(a). In Fig.\ \ref{hall_cond_topological.fig} (main) we plot $\sigma_{xy}$ as a function of $m$ for various temperatures. Let us first focus on the low temperature case $T=0$ K (blue, solid line), where we assume the Fermi level to lie at $E_F=0$ meV in the middle of the VB and CB. The stored bit can be read out simply by measuring $V_H$ and determining its sign. In this case $\sigma_{xy}$ is half-quantized at $e^2/2h$ for $m\neq 0$ and is topologically robust to cell imperfections. For a typical driving current of $I=1$ $\mu$A,\cite{klitzing} this corresponds to a readout voltage of $V_H=\pm47$ mV.

At finite temperatures $T>0$ K, the Fermi distribution of the carriers must be factored into the calculation of $\sigma_{xy}$. Moreover, the size of the gap is important as the bulk insulating behavior can be destroyed by thermal excitations from the VB to CB (we require $\Delta\gg k_B T$). Increasing the gap size whilst maintaining $E_F$ to lie within the gap may be achievable through the doping technique outlined \emph{e.g.}\ in Ref.\ \onlinecite{ylchen}. There, it was found that doping Bi$_2$Se$_3$ with Fe resulted in an opening of a gap together with an upward shift of the Fermi level into the CB (making it $n$-type). The Fermi level could then be re-shifted back into the gap by introducing non-magnetic $p$-type dopants. Refining this two-step procedure of (i) opening the gap, and (ii) shifting the Fermi level into the induced gap could potentially accommodate very large gaps, whilst maintaining the Fermi level to lie inside the gap. Fig.\ \ref{hall_cond_topological.fig} (main) shows the effect of increasing $T$ well beyond $0$ K. In our calculations, we assumed that the Fermi level lies at $E_F=0$ meV, \emph{i.e.}\ always within the gap. Fig.\ \ref{hall_cond_topological.fig} (main) shows that the thermal effect diminishes $\sigma_{xy}$ from its quantized value at $T=0$ K, but that the accuracy is improved with increasing $|m|$. A large $|m|$ is also beneficial as it helps to preserve bulk insulating behavior as discussed above. For illustration, we indicate the points $m=k_B T$ for each $T>0$ K (corresponding to $\Delta = 2k_B T$). From a mean field perspective, the exchange splitting $\Delta=2|m|$ is given by $\Delta=nJ\langle S\rangle$, \cite{ryu} where $n$ is the doping concentration, $J$ is the exchange coupling and $\langle S\rangle$ is the expectation of the local spin at saturation. Using typical values of $J=50$ meVnm$^2$, \cite{garate} and $\langle S\rangle=1.5$ $\mu_B$ for Mn-doped Bi$_2$Te$_3$,\cite{yshor} a value of $m=30$ meV corresponds to a doping concentration of $n=0.8$ nm$^{-2}$ which is of the order of typical values.\cite{prl106_136802} In Fig.\ \ref{hall_cond_topological.fig} (inset) we study the effect of $E_F\neq 0$ for $T=100$ K (for illustration see Fig.\ \ref{device.fig}(c)), which indicates a general broadening effect. The sloped regions coincide with the condition $|m|<E_F$, where the Fermi level lies inside the CB. In our device, it is desirable to ensure that $|m|>E_F$, such that $\sigma_{xy}$ is quantized (apart from scaling by the Fermi distribution). Greater accuracy of $\sigma_{xy}$ is achieved for $|m|\gg E_F$.
%

A topologically invariant readout process is attractive from the point of view of robustness to impurities and geometrical imperfections, such as edge roughness, in analogy with the IQHE. It also alleviates the use of voltage comparators which traditionally compare the readout signal to a threshold voltage to determine stored bits; such processes are prone to noise which may lead to errors in bit detection. Moreover, once $|m|$ is sufficiently large, $\sigma_{xy}$ exhibits only a weak dependence on $m$. In practice, the writing process is not perfect; $m$ is not fully switched to the vertical direction and will exhibit spatial fluctuations. Our proposed memory cell ensures a constant readout voltage even in the presence of such imperfections. 

Despite the advantages for a topological memory cell or magnetic sensor, several challenges are anticipated, such as opening up a sufficiently large gap for high temperature operation (achieving large $|m|$). Furthermore, the Curie temperature $T_C$ of magnetic TI surfaces needs to be improved drastically. Presently, experiments reveal that $T_C \lesssim 20$ K for Mn\cite{yshor} and Fe\cite{kul} doped Bi$_2$Te$_3$. However, there are hopes of increasing $T_C$ beyond $100$ K \cite{ryu} in the same spirit as magnetic III-V semiconductors,\cite{ohno} for the mechanism for magnetism in the two systems are analogous.\cite{yshor}

In summary, we have proposed a memory cell based on magnetically doped topological insulators. Writing information to the cell entails switching the cell magnetization. The readout process is facilitated by the Hall effect, which is a function of the stored information. The Hall voltage is a topological quantity which is insensitive to details such as edge roughness, the presence of impurities and defects, and imperfect writing. 
%

\end{document}